# Accelerated MR Elastography Using Learned Neural Network Representation

Xi Peng

Affiliations:

Departments of Radiology, Biomedical Engineering, Electronic and Computer Engineering, and Mechanical Engineering, University of Iowa, Iowa City, IA 52246

**Running title:** MRE Image Reconstruction with Learned Network Representation

**Correspondence:**
**Xi Peng, Ph.D.**
Department of Radiology
University of Iowa, Iowa City, IA, 52246
Email: xi-peng@uiowa.edu; stevep1120@gmail.com;

# ABSTRACT

**Purpose:** To develop a deep-learning method for achieving fast high-resolution MR elastography from highly undersampled data without the need of high-quality training dataset.

**Methods:** We first framed the deep neural network representation as a nonlinear extension of the linear subspace model, then used it to represent and reconstruct MRE image repetitions from undersampled k-space data. The network weights were learned using a multi-level k-space consistent loss in a self-supervised manner. To further enhance reconstruction quality, phase-contrast specific magnitude and phase priors were incorporated, including the similarity of anatomical structures and smoothness of wave-induced harmonic displacement. Experiments were conducted using both 3D gradient-echo spiral and multi-slice spin-echo spiral MRE datasets.

**Results:** Compared to the conventional linear subspace-based approaches, the nonlinear network representation method was able to produce superior image reconstruction with suppressed noise and artifacts from a single in-plane spiral arm per MRE repetition (e.g., total R=10), yielding comparable stiffness estimation to the fully sampled data at 2mm resolution.

**Conclusion:** This work demonstrated the feasibility of using deep network representations to model and reconstruct MRE images from highly-undersampled data, a nonlinear extension of the subspace-based approach.

# 1. INTRODUCTION

Brain magnetic resonance elastography (MRE) is a powerful tool to noninvasively access tissue biomechanical properties in vivo, with broad potential in various neuroimaging applications including normal pressure hydrocephalus[1], brain tumor[2], neurodegenerative disease[3], cognitive function[4] and brain development[5]. Higher spatial resolution is particularly important for identifying heterogenous substances and accurately quantifying stiffer tissues. Achieving high-resolution MRE relies on two key factors: fast imaging and robust stiffness inversion. This paper focuses on advancing the fast-imaging component using deep learning-based reconstruction.

MRE is a phase-contrast MRI technique and typically requires the acquisition of multiple imaging repetitions for encoding wave propagations in the 3D vector space (e.g., 6 directions) at varying vibration phase-offsets. As a result, achieving high spatial resolution within a practical scan time is challenging. To address this issue, fast scan such as echo-planer-imaging (EPI) and spiral[6], along with more efficient motion-encoding strategies have been developed, including SLIM[7] (sample interval modulation), Ristretto[8], DENSE[9, 10] (multiphase Displacement ENcoded Stimulated Echo), Magnetization-Prepared[11], and distributed encoding MRE[12]. On the other hand, constrained reconstruction from undersampled data has been extensively exploited for fast imaging, with compressed sensing[13, 14] and low-rank modeling[15, 16] being among the most widely adopted approaches leveraging the spatial and temporal redundancies. Additionally, anatomical and phase constraints are among the "early" constrained methods used in various MR applications. For instance, the anatomical similarity of a series of MR images has been utilized for denoising and joint image reconstruction via spatially-adaptive regularization[17, 18], generalized series model[19, 20], group sparsity[21, 22] and Bayesian modeling[23]; Phase constraints imposing smoothness have been exploited in standard partial Fourier parallel imaging and phase-contrast applications via imaginary component regularization[24, 25], structured low-rank approximation[26] and direct spatial regularization of the phase with separate magnitude and phase modeling. [27-29]

Recently, deep learning (DL)-based reconstructions have shown considerable promise in producing high quality MR images from under sampled data. However, DL-based reconstruction for MRE is relatively underdeveloped probably because 3D high-resolution high-quality MRE datasets are difficult to acquire for training supervised models[30]. That said, unsupervised learning-based reconstruction appears to be a natural option for MRE. For example, end-to-end unrolled networks trained using zero-shot self-supervised learning has been proposed for knee imaging[31], cardiac imaging[32] and quantitative imaging[33]; deep image prior (DIP) or manifold representation using a deep neural network have also been successfully developed for cardiac imaging[34, 35], quantitative imaging[36], upper airway imaging[37], and free-breathing pulmonary imaging[38].

In this work, we first establish a direct connection between deep neural network representation and linear subspace models[39], viewing the former as a nonlinear extension of the latter; then we propose to use the neural network representation to reconstruct MRE images from highly undersampled data. Additional elastography-specific priors were further incorporated to improve reconstruction quality. Experiments on both multi-slice spin-echo and 3D gradient-echo spiral MRE datasets were conducted to demonstrate the feasibility of the proposed method.

# 2. METHODS

## 2.1 From Low-rank to Deep Neural Network Representation

According to the partial separability theory[16], a spatiotemporal function $\rho(\boldsymbol{r}, t) \in \mathbb{C}^{N \times T}$ in MRI can usually be approximately by a linear subspace model, whose matrix form can be expressed as:

$$\rho = UV^H$$

where $U \in \mathbb{C}^{N \times L}$ and $V \in \mathbb{C}^{T \times L}$ are the spatial coefficients and temporal basis, respectively. $N$ is the total number of voxels, $T$ is the total temporal frames/repetitions, $L$ is the model order or the rank. At each time point $t$, the image function $\rho_t$ can be rewritten as:

$$\rho_t = Uv_t^H$$

where $v_t \in \mathbb{C}^{1 \times L}$ denotes the row vector of $V$ corresponding to time point $t$. In other words, each image voxel in $\rho_t$ can be considered as a linear combination of elements in $v_t$ with different weights stored in $U$ (i.e., row vectors of $U$). Note that, the low-rank model inherently assumes that image voxels from different time points share the same weights in $U$. We further observed that this linear representation is equivalent to a fully connected layer with identity activation function as illustrated in Fig. 1a. Grounded on this reformulation, the deep network representation can be considered as a nonlinear extension of the linear subspace model by simply replacing the fully connected layer with a deep neural network:

$$\rho_t = G(v_t|\theta)$$

where $G(\cdot)$ is an image generative network or decoder with trainable parameter $\theta$, which creates a feature-based representation of the spatial variations. The temporal variations are also assumed to reside in a low-dimensional manifold represented by the latent vectors, which can be initially derived from measurement data. We note that, implementation-wise, the proposed neural network representation is very similar to the existing Time-DIP[34] or manifold learning methods[35, 37] where they initialize $v_t$ with random Gaussian noise. Direct comparison was provided in the Results section.

## 2.2 MRE Image Reconstruction with Deep Network Representation

Given $\rho_t = G(v_t|\theta)$, the reconstruction is equivalent to estimating θ and $v_t$ from the measured data. Similar to existing self-supervised approaches[31, 37], by substituting the network representation into the forward imaging equation, the cost function for MRE image reconstruction can be written as:

$$\underset{\theta, v_t}{\operatorname{argmin}} \sum_{t=1}^{T} \sum_{i=1}^{C} ||d_{t,i} - FS_i\rho_t||_2^2 + \lambda_1 R_{magn}(\boldsymbol{\rho}) + \lambda_2 R_{wave}(\boldsymbol{\rho}) + \lambda_3 R_v(\boldsymbol{v}), \quad (2)$$

where $d_{t,i}$ denotes the k-space data from the *t*-th repetition and *i*-th coil. $S_i$ is the coil sensitivity function. Initialized from the temporal basis, the latent vectors $\boldsymbol{v} = \{ v_1, v_2, \ldots, v_T \}$ enables the nonlinear neural network to generate corresponding temporal images through its powerful and flexible spatial representation capability. To further enhance reconstruction quality, additional regularizations (e.g., $R_{magn}$ and $R_{wave}$) are introduced to impose anatomical similarity among image repetitions and smoothness of the wave-induced displacement. The regularization parameters $\lambda_1$, $\lambda_2$ and $\lambda_3$ are set a priori in this study. The network parameters θ and $v_t$ are jointly trained and updated using the above minimization problem in a self-supervised manner. More details are provided below.

### 2.2.1 Multi-level data consistency

In this work, we used a standard multi-level decoder network to represent the MRE images (Fig. 2). Specifically, similar to the Time-DIP[34] method, a three-layer perceptron was first employed to replace the linear fully connected layer in the subspace model, mapping $v_t$ to a feature space with dimension $Q = N/K^2$. $K$ is the total number of levels or scales in the network. Afterwards, a convolutional decoder network was exploited to restore the full resolution image $\rho_{t,K} \in \mathbb{C}^{\sqrt{N}\times\sqrt{N}}$ from the coarsest features $f_{t,1} \in \mathbb{C}^{\sqrt{Q}\times\sqrt{Q}}$. At each level of the decoder, a linear interpolation operator was used to raise the size of feature maps by a factor of 2, followed by a residual CNN block to learn the higher resolution features. To better guide the updating of network weights in shallow layers, we adopted the idea of multi-level deep supervision from an early 3D U-net segmentation method[40] (i.e., data-consistent losses in our case) that are naturally suited to spiral MRE data. More specifically, at the end of each level, feature maps $f_{t,k}$ were combined into intermediate images $\rho_{t,k}$ by a single convolutional layer, followed by additional data-consistent losses calculated using corresponding k-space spiral segments $d_{t,i,k}$:

$$\underset{\theta, v_t}{\operatorname{argmin}} \sum_{t=1}^{T} \sum_{i=1}^{C} \sum_{k=1}^{K} ||d_{t,i,k} - F_k S_{i,k} \rho_{t,k}||_2^2 + \lambda_3 R_v(\boldsymbol{v})$$

where $\rho_{t,k}$ denotes the intermediate image output at the $k$-th resolution level for repetition $t$, i.e., $G(v_t|\theta) = \{\rho_{t,1}, \rho_{t,2}, \dots, \rho_{t,K}\}$. $F_k$ and $S_{i,k}$ are the corresponding Fourier transform and coil sensitivity for the $k$-th resolution. $R_v(\cdot)$ is a regularization imposed on latent vector $\boldsymbol{v}$, simply selected to constrain its energy:

$$R_v(\boldsymbol{v}) = \|\boldsymbol{v}\|_2^2$$

### 2.2.2 Magnitude similarity constraint

The magnitude signal capturing the anatomical structures in MRE data remains similar among repetitions. Although a single decoder network is used to model MRE images across different repetitions, the anatomical structures among repetitions might still be able to vary due to the high nonlinearity of the neural network. Therefore, additional constraint is needed to enforce similar anatomical structures among repetitions. Previously, anatomical constraints have been developed for diffusion-weighted[18] imaging and MRE[41] respectively by joint adaptive spatial regularization using reweighted L2 norms. Note that those regularizations were imposed on the complex-valued image quantities, affecting both magnitude and phase information. In this work, we propose a different anatomical similarity constraint mainly applied to the magnitude signal at each scale $k$:

$$R_{magn}(\boldsymbol{\rho}) = \sum_{i,j \in T, i \neq j} \sum_{k=1}^{K} \left\| |\rho_{i,k}| - |\rho_{j,k}| \right\|_1$$

where $|\cdot|$ takes the magnitude of a complex quantity, $\|\cdot\|_1$ denotes the $l_1$ norm. Notably, by minimizing the $l_1$ norm of magnitude differences, magnitude signals are expected to be close but not strictly the same to allow for the potential intravoxel phase disperse[42]. This is also a softer magnitude regularization compared to the separate magnitude and phase models[27, 29].

### 2.2.3 Wave-induced displacement smoothness

The vibration displacement is encoded in the phase of the MRE images, along with the static image phase. In realistic viscoelastic materials undergoing micron-level harmonic vibrations, such as the brain, tissue displacement can be considered spatially smooth. On the contrary, in gradient-echo acquisitions, the static image phase may contain high-frequency components due to tissue susceptibility-induced field variations. Therefore, in this work, we propose to promote smoothness only on the vibration displacement using a standard spatial regularization[29]:

$$R_{wave}(\boldsymbol{\rho}) = TV(H\boldsymbol{\rho})$$

where $TV(\cdot)$ represents the total variation operator. Operation $H$ computes the exponential term of the wave-induced phase components ($\varphi_t$), i.e., $H\boldsymbol{\rho} = \{e^{i\varphi_t} | t = 1,2, \dots, T\}$ with $\rho_{\boldsymbol{t}} = \rho_{\boldsymbol{0}} e^{i(\varphi_0+\varphi_t)}$, for each motion encoding direction and phase offset. Note that imposing smoothness constraint on the wave-induced phase will not compromise the static anatomical and tissue susceptibility information ($i.e., \rho_{\boldsymbol{0}} e^{i\varphi_0}$).

## 2.3 Experiments

In vivo experiments were performed on a 3T MR scanner using a 14-channel receive-only head coil. All experimental data were obtained with the approval of institutional review board and written informed consent was provided from healthy subjects. Mechanical vibrations at 60 Hz were introduced into the subject's head along anterior–posterior using a commercial pneumatic active driver (Resoundant, Rochester, MN) and a cushion-like passive driver placed underneath subjects' head. Four wave-propagation phase offsets were sampled evenly over a single wave period by adjusting the synchronization between the sequence and active driver. Six motion encoding gradients (MEG) along positive and negative x, y, z directions were applied repeatedly to acquire the 3D displacement field. All MRE dataset in this study share the same brain coverage (FOV = 240×240×120 $mm^3$), MEG amplitude and driver vibration strength.

Both fully sampled multi-slice spin-echo spiral MRE dataset (TR/TE = 5000/50 ms, flip angle = 90°, scan time = 10 minutes, 13.5 ms spiral readout, 5 spiral arms, 2mm isotropic resolution) and in-plane fully sampled 3D gradient-echo spiral staircase MRE dataset[41] (TR/TE = 66.67/35 ms, FA = 23°, 1.2× slice oversampling, 2× through-plane uniform undersampling, scan time = 5 minutes, 13.6 ms spiral readout, 5 spiral arms, 2mm isotropic resolution) were obtained. To evaluate the performance of the proposed reconstruction, the data were then retrospectively undersampled to a single spiral arm per repetition interleaving across all repetitions (i.e., each repetition used a different spiral arm), mimicking a combined in-plane and temporal acceleration factor of 5. (i.e., a total acceleration factor of 10 if considering the existing 2× through-plane undersampling of the 3D GRE data, namely only 1 min scan time for 2mm isotropic resolution). Standard SENSE and low-rank reconstructions (i.e., rank = 12) were conducted for comparison. The relative L2 norm errors were computed to quantitatively access the quality of reconstructed anatomical and wave images, i.e., $\|\hat{\rho} - \rho_0\|_2^2 / \|\rho_0\|_2^2$, where $\hat{\rho}$ is the reconstructed image and $\rho_0$ is from the fully sampled data. To further compare performance of the low-rank and the proposed methods, reconstructions were conducted across different acceleration factors (i.e., varying number of undersampled arms). To quantitatively access the estimated stiffness maps, both the median stiffness values and voxel-wise stiffness estimation errors as compared to the fully sampled data were reported in gray and white matters. Finally, to evaluate the effect of each component in the proposed method, ablation studies were conducted using both the spin-echo and 3D GRE SSC datasets. The Time-DIP[34] method was directly compared with the proposed subspace initialization followed by the addition of multi-level data consistency loss and elastography-specific constraints. The distinct effects of magnitude similarity and displacement smoothness constraints were studied using varying levels of regularization parameters.

## 2.4 Implementation Details

### 2.4.1 Formulation of 3D spiral staircase (SSC) Reconstruction

In contrast to stack-of-spiral, spiral staircase trajectory uniformly distributes spiral arms of different orientations across the Nyquist interval $\Delta k_z$. Notably, the shift in $k_z$ between different orientation groups produce a linear phase variation along slice in the image domain. The additional phase variation introduces more incoherence that is beneficial for through-plane parallel imaging. Such linear phase term can be pre-calculated based on the designed trajectory and needs to be incorporated in the forward imaging model. More details can be found in a previous work[41]. In this paper, the reconstruction of 3D SSC data was decoupled at each aliased slice location (i.e., index z), similar to the unfold-SENSE formulation:

$$\widehat{\boldsymbol{\rho}}_z = \underset{\widehat{\boldsymbol{\rho}}_z}{\operatorname{argmin}} \sum_{c=1}^{L} \sum_{a=1}^{N_a} \left\| \boldsymbol{F}(\boldsymbol{P}_{a,z} \odot \boldsymbol{S}_c) \boldsymbol{\rho}_z - \boldsymbol{d}_{c,a,z}^{aliased} \right\|_2^2$$

where $\boldsymbol{\rho}_z \in \mathbb{C}^{2N\times 1}$ denotes the superimposed images at slice location z1 and z2 (e.g., $\boldsymbol{\rho}_z$=$[\boldsymbol{\rho}_{z1}, \boldsymbol{\rho}_{z2}]^{\mathrm{T}}$), $\boldsymbol{S}_c = [\boldsymbol{S}_{c,z1}, \boldsymbol{S}_{c,z2}]$ are the corresponding coil sensitivities, $\boldsymbol{P}_{a,z} = [\boldsymbol{P}_{a,z1}, \boldsymbol{P}_{a,z2}]$ represents the linear phase variation at each slice location for spiral arm $a$, $\boldsymbol{d}_{c,a,z}^{aliased}$ is the slice-aliased data obtained by applying inverse Fourier transform along $k_z$ to the undersampled k-space, $\boldsymbol{d}_{c,a,z}^{aliased} = [\boldsymbol{F}_z^{-1}\boldsymbol{d}]_{c,a,z}$. Network formulation for the 3D SSC dataset was further descried below.

### 2.4.2 Network training

The multi-level network was constructed in Pytorch and trained with all MRE repetitions as a single batch (e.g., a batch size of 24 consisting data from 4 phase offsets and 6 directions) using a gradient descent algorithm with an Adam optimizer on one Nvidia A100 80GB GPU for a fixed number of iterations (500 iterations for the 2D spin-echo dataset and 1000 iterations for the 3D GRE dataset). Complex-valued CNN and activation functions were adopted due to their superior performance in phase-based applications[43, 44].

To reduce GPU memory cost, training was performed on a "slice-by-slice" basis. Initial input vectors $v_t$ were computed using singular value decomposition (SVD) from the center of measured k-space, similar to the standard subspace-based methods[15, 16]. For the 3D GRE dataset, $\boldsymbol{d}_z^{aliased}$ was used in the data consistency loss for the z-th aliased slice. Compared to the multi-slice case, the structure and capacity of MLP and ResNet remain unchanged except that 1) two feature maps and image slices were produced by the MLP and ResNet at the output layers; 2) $v_t$ was formed by concatenating subspaces derived from $\boldsymbol{P}_{z1}^{H}\boldsymbol{d}_z^{aliased}$ and $\boldsymbol{P}_{z2}^{H}\boldsymbol{d}_z^{aliased}$, which was also used in the low-rank method. Since the neural network was used to produce two slices simultaneously in this case, the total number of iterations was doubled in 3D as compared to the 2D case.

The Sigpy package[45] was adopted to perform non-uniform FFT for spiral image reconstruction (i.e., explicitly constructed forward and adjoint NUFFT operators). Off-resonance correction was conducted after image reconstruction using a deblurring-based method[46]. The coil-sensitivity and field inhomogeneity maps were obtained separately using a vendor-provided pre-scan. Coil sensitivity maps were then interpolated at different resolution levels for computing the multi-level data consistent loss.

### 2.4.3 Shear modulus inversion

After obtaining the deblurred MRE images, wave-induced displacements were calculated by phase subtraction of the image pairs with positive and negative motion encoding gradient (MEG). Low frequency component (e.g., 3×3) of the displacement field was first removed to reduce bulk motion. Afterwards, a Laplacian based algorithm[47] was used to unwrap the displacement field, followed by a denoising step using the divergence-free wavelet transformation.[48] The complex displacement vector fields were then extracted from the 1st harmonic component using temporal Fourier transform. Finally, the displacement fields were further smoothed by a 3D 4th order Butterworth lowpass filter[49] with a threshold of 100 $m^{-1}$ before using the algebraic inversion[50] to produce the complex share modulus ($G^* = G' + iG''$). Shear stiffness was then computed based on: $\mu_{stiff} = 2|G^*|^2/(G' + |G^*|)$. To quantify the estimated stiffness, the median values and voxel-wise stiffness errors were reported within gray matter and white matter respectively, whose masks were segmented from anatomical images using FSL (FMRIB Software Library). The stiffness maps were filtered with a 3×3×3 median filter for visualization.

## 3. RESULTS

The reconstructed MRE images, displacement and estimated shear stiffness maps (i.e., axial, sagittal and coronal views) from an undersampled 3D GRE and a multi-slice SE dataset are shown in Fig. 3 and Fig. 4 respectively, as well as in supplementary Fig. S1 and Fig. S2. As can be seen, conventional SENSE reconstruction suffered from severe aliasing artifacts. The state-of-the-art low-rank reconstruction also exhibited noticeable aliasing artifacts at this acceleration, leading to degraded displacement field and highly underestimated stiffness maps. On the contrary, the network representation method was able to produce high-quality MRE images with better delineated anatomical details and improved SNR as well as comparable displacement and stiffness estimation to the fully sampled data. Similar results were observed in both GRE and SE datasets.

Results of reconstructions with varying number of spirals arms were illustrated in Fig. S3. As expected, the low-rank approach was able to produce good-quality MRE images when downsampled to two spiral arms (i.e., from five fully sampled ones) but its performance degraded when undersampling is further increased to a single spiral arm. This is because the inherent degree of freedom in the MRE dataset with current 24 repetitions is 12 for the linear subspace model[15], namely it only allows a temporal acceleration of less than 2. However, this threshold was violated when using only 1 spiral arm across different repetitions. On the other hand, the network representation approach (i.e., without the magnitude similarity and displacement smoothness constraint) consistently generate good-quality MRE images across all acceleration cases, as manifested by the estimated stiffness maps.

Quantitative evaluations of the reconstructed MRE image and estimated stiffness maps were reported in Fig. 5 and Fig. 6. Specifically, the median and range of estimated stiffness values in gray and white matters are reported in Fig. 5a and Fig. 5b. The relative L2 norm errors of the reconstructed images from varying number of arms are summarized in Fig. 5c. With a single spiral arm, the proposed method produced stiffness measurements comparable to the fully sampled data, while standard SENSE and low-rank methods significantly underestimated shear stiffness due to severe aliasing artifacts and poor SNR (Fig. 5a). Additionally, as the number of spiral arms decreased from 5 to 1, the network method consistently produced reasonable image reconstruction and stiffness estimates (i.e., though quantification gradually decreased due to compromise in SNR), while the performance of the low rank method dropped significantly (Fig. 5b and 5c). Figure 6 shows the quantitative comparison of the stiffness estimation errors using the spin-echo dataset. The stiffness quantification accuracy using the low-rank method is significantly biased and skewed while the proposed method provides nearly "unbiased" mean stiffness values with symmetrically distributed random errors, again demonstrating its superiority to the low-rank method.

Results of ablation study was displayed in Fig. 7 and Fig. S4. Figure. 7 shows that random initialization and subspace initialization yield very comparable anatomical image and wave reconstructions (although subspace initialization appears slightly better quantitatively). Incorporating multi-level DC loss reduces overall noise and artifacts, and the proposed elastography-specific constraints further improves the anatomical details and wave propagation. The individual effectiveness of magnitude similarity and displacement smoothness constraints are further demonstrated in Fig. S4. Both qualitatively and quantitatively, they lead to additional suppression of artifacts in the anatomical and wave images. However, the resulting stiffness maps appear largely similar, likely due to the strong smoothing operation required in the direct inversion process.

Lastly, the distinct roles of magnitude and displacement constraints are illustrated in Fig. 8, where anatomical and wave images are shown respectively under varying levels of regularizations (i.e., weak, moderate and strong). As can be seen, each regularization primarily affects its respective domain. Specifically, increasing the magnitude similarity regularization progressively smooths the anatomical image, while the wave image remains largely unchanged with a fixed displacement regularization. This magnitude similarity constraint effectively introduces an averaging effect across MRE repetitions. Similarly, increasing the displacement regularization yields progressively smoother wave images, with minimal impact on the anatomical image under a fixed magnitude regularization. However, excessive regularization in either domain results in over smoothing of either magnitude or displacement images. These complementary effects further underscore the effectiveness of the proposed regularization strategies.

# 4. DISCUSSION

The main contribution of this work is that we conceptually framed the deep neural network representation as a nonlinear extension of the linear subspace model and proposed to use it for MRE image reconstruction for the first time. From an implementation standpoint, we initialize the latent vectors using the linear temporal basis rather than random Gaussian noises in the exiting manifold learning approaches[34, 35, 37]. Similar ideas have been exploited for standard DIP-based approaches, such as self-guided DIP[51], reference-guided DIP[52], where a reference or an intermediate image is adopted as input to the network. The underlying assumption is that it would be easier to map the target image function from certain image/data-specific space than from random Gaussian noise. The ablation study showed that the temporal subspace initialization yields reconstruction that are qualitatively comparable to that from random initialization, with only marginal quantitative improvement. This is probably because temporal variations are relatively easier for the network to capture than spatial variations. Our second contribution is the introduction of multi-level data consistency and elastography-specific loss terms, which further suppresses noise/artifacts and preserves image details.  Although the gains in anatomical detail preservation and wave artifacts suppression provided by the regularization terms are diminished in the conventional direct inversion process for MRE, they could be beneficial for more advanced inversion approaches, such as non-linear inversion and learning-based inversion, as well as for applications involving simultaneous MRE and QSM[53], where the static high-resolution local tissue field is essential.

Compared to the unrolled self-supervised network methods[31, 32], we used the neural network to represent the spatiotemporal image instead of learning an end-to-end mapping from undersampled to high-quality images. A couple of advantages can be highlighted. First, the training strategy is simpler. We do not need to randomly reallocate the undersampled data to the loss function and data consistency layer each time. Second, the network used for image representation is typically much more lightweight than the network for end-to-end mapping. Both factors contribute to more efficient training as demonstrated in a recent study comparing network representation with End-to-End unrolled network for quantitative T2 mapping[54]. Another promising nonlinear representation for image reconstruction is the implicit neural representation[55] (INR). The key difference is that our method can be viewed as a nonlinear extension of the linear subspace model, whereas INR may be interpreted as a nonlinear extension of the finite-dimensional linear model (with discrete voxel functions on Cartesian grids). As such, INR represents a promising alternative to the proposed approach and warrants further investigation.

We demonstrated that the network representation approach could produce high-quality MRE images from a single spiral arm in the in-plane k-space at 2mm resolution while the performance of state-of-the-art low-rank method degrades substantially. Additional MRE-specific constraints, including magnitude similarity and displacement smoothness, were further incorporated to suppress noise and artifacts in the magnitude and displacement field. Generally, the proposed method brings the previous 3D GRE spiral staircase acquisition down to 1 minute at 2mm isotropic resolution. Although this study primarily used spin-echo and gradient-echo spiral datasets, the approach is adaptable to all sequence types and sampling trajectories. We believe that the proposed method is particularly useful for high-resolution MRE when both higher acceleration and reasonable scan time are desired. Note that, the representation capability of the network is governed by its architecture, which may selectively suppress or preserve certain desirable or undesirable images features. For instance, in the 3D GRE data, the “whirling-like” artifacts (Fig. S3) were less represented by the neural network even when using fully sampled dataset. This artifact is most likely caused by the pronounced T2* decay near the nasal cavity, which accumulates along the spiral readout (16 ms), particularly given the relatively long TE (35ms) used in this GRE sequence. An implementation with stronger MEG, thus shorter MEG duration and shorter TE (e.g., 19ms) would help mitigate this artifact. To prevent over fitting, we simply fixed the total number of iterations for the 2D and 3D cases, respectively, as a form of early stopping. The proposed multi-level data consistency loss appears to further mitigate overfitting by emphasizing more on the low-frequency k-space (Fig. 7). In addition, self-validation scheme could also be adopted.[56]

The current work has several limitations. First, physiological errors in the presence of MEGs are potential confounding factors especially for multi-shot spirals which have not been explicitly addressed. With variable density spirals, the oversampled center can be used as self-navigator to mitigate the bulk phases offsets and k-space shifts as proposed in the literature[57]. We empirically observed that 3D GRE-based MRE with whole-brain excitation seems relatively more robust to bulk motion especially along the slice direction, probably because it smears the phase errors across the entire 3D k-space. Results from another healthy subject obtained using the 3D GRE-SSC sequence are provided in Fig. S5; Second, the performance results were demonstrated on only one dateset at a time. Additional validation on more subjects' data is needed in the future; Third, conventional algebraic inversion was utilized to derive the stiffness maps which is quite sensitive to noise or any local image artifacts. Both fully sampled and undersampled reconstructions yield a few artificially elevated stiffness values due to the direct inversion process. Advanced inversion method is the other important pilar for achieving high resolution MRE which is out of the scope of this paper; and lastly, the model was trained from scratch using a single undersampled dataset, thus requiring relatively long computational time. State-of-the-art supervised learning was not fully exploited for MRE due to the lack of ground truth data, but synthesized data might be useful to pretrain the neural networks for MRE image reconstruction. Future work will include incorporating both synthesized data and multiple undersampled data for pretraining and to reduce computational time at inference.

## 5. CONCLUSION

We presented a self-supervised learning method for improved MRE image reconstruction from highly undersampled data, which can be considered as a nonlinear extension of the linear subspace-based approaches conceptually. We believe this work will serve as an important pilar for advancing high-resolution brain MRE.

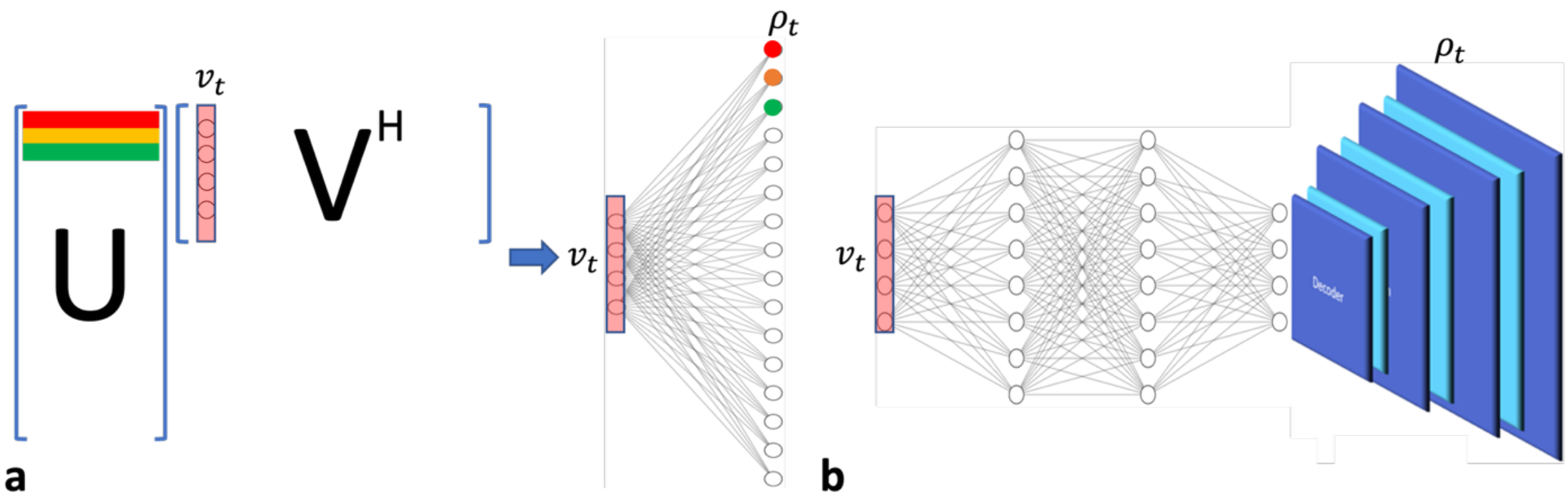

Figure 1. Illustration of a) the linear subspace model and b) the deep neural network representation as the nonlinear extension.

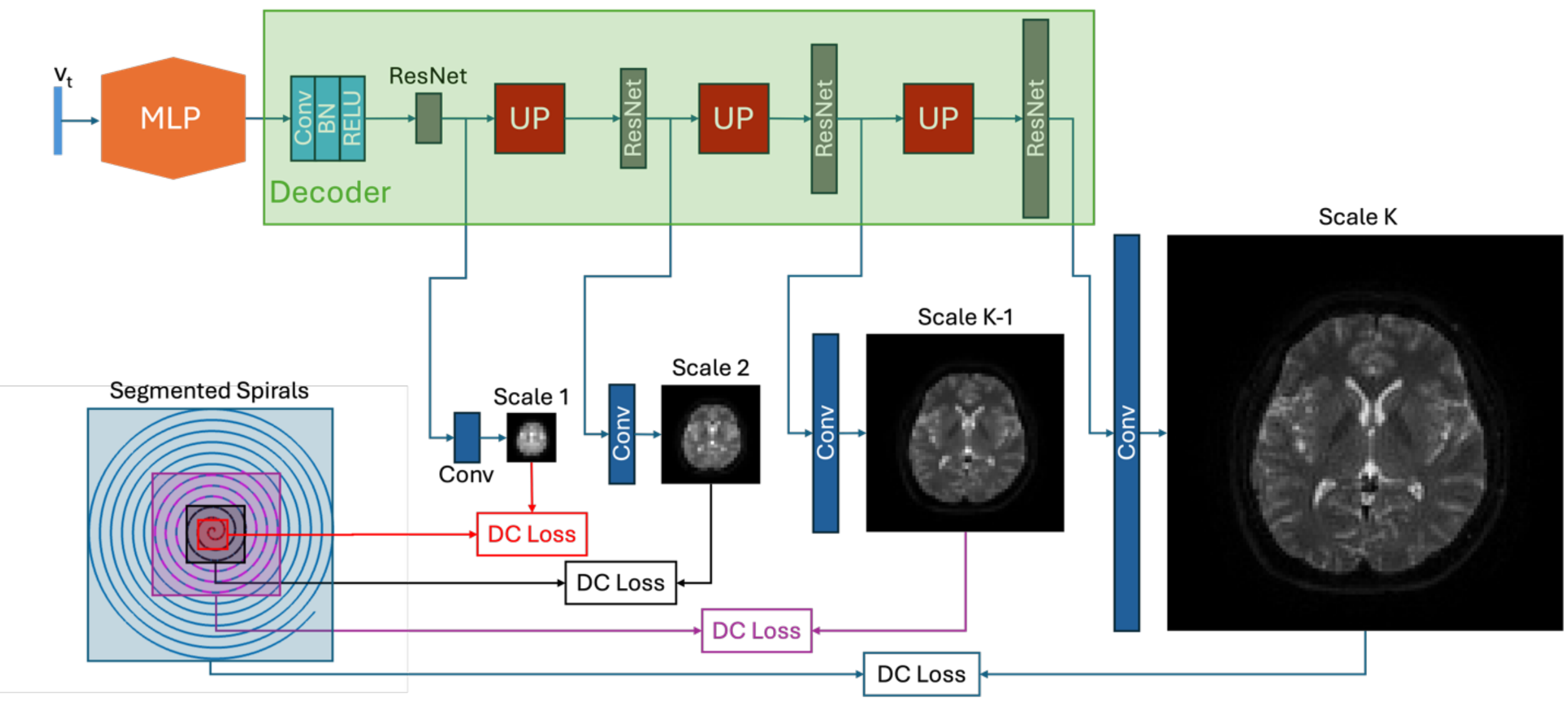

Figure 2. The multi-level architecture of the neural network decoder with multi-level k-space consistent losses.

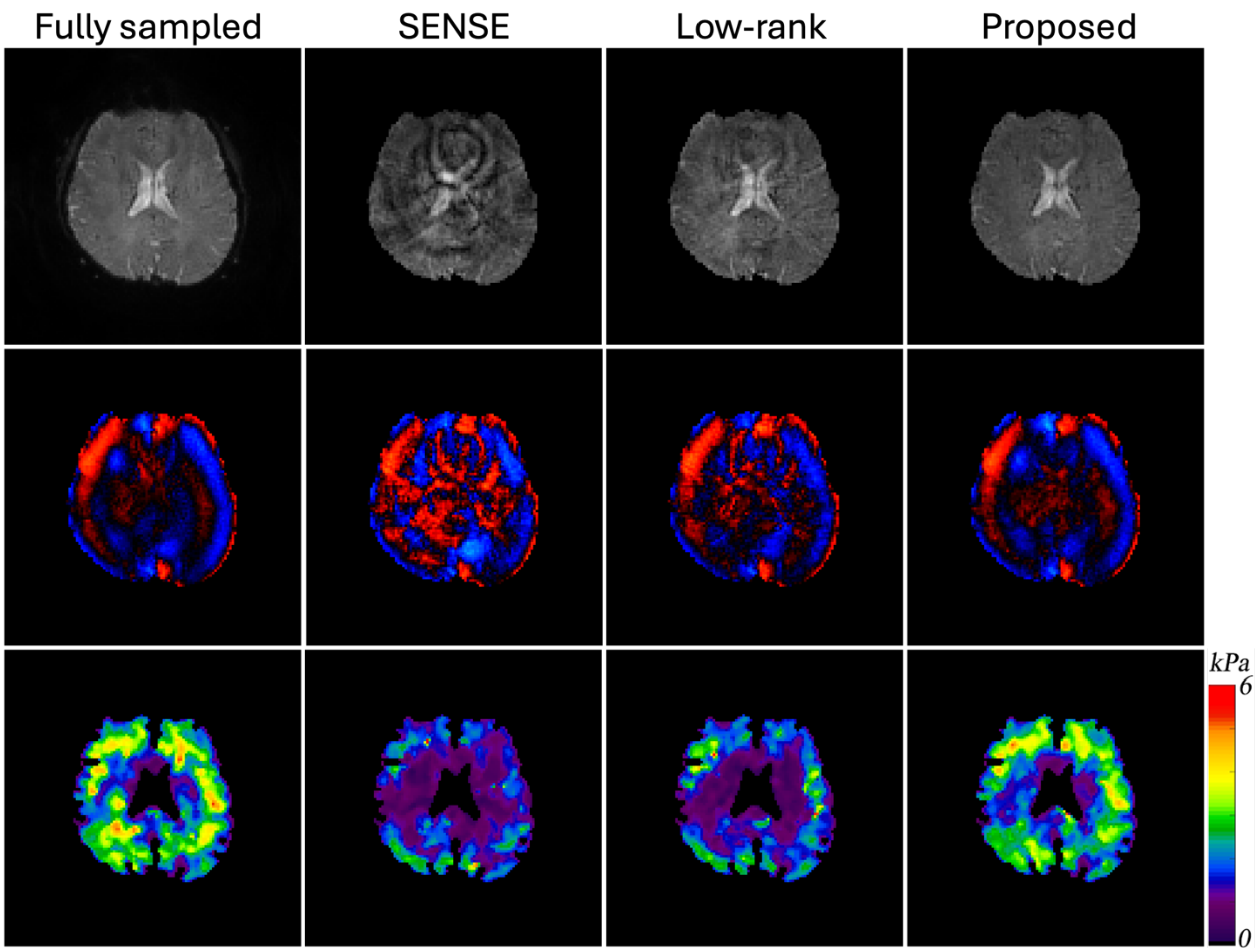

Figure 3. Reconstructed MRE images, displacement (along Y direction) and stiffness maps using a single spiral arm (out of the fully sampled 5 spirals) from the 3D GRE-SSC dataset (i.e., a total reduction factor of 10 if accounting for the 2× through-plane acceleration).

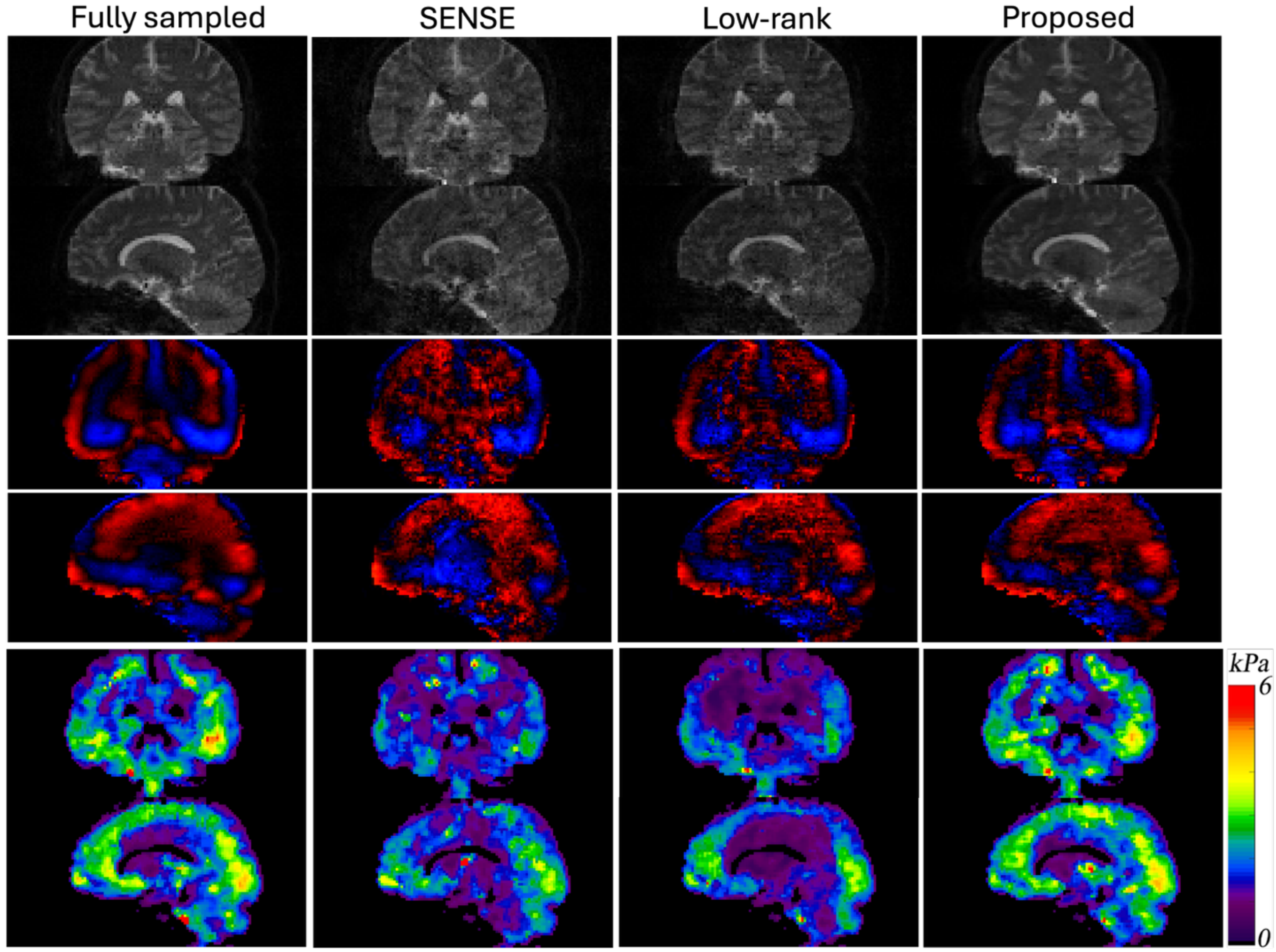

Figure 4. Reconstructed MRE images, displacement (along Y direction) and stiffness maps using only 1 spiral arm (out of fully sampled 5 spirals) from the multi-slice spin-echo dataset.

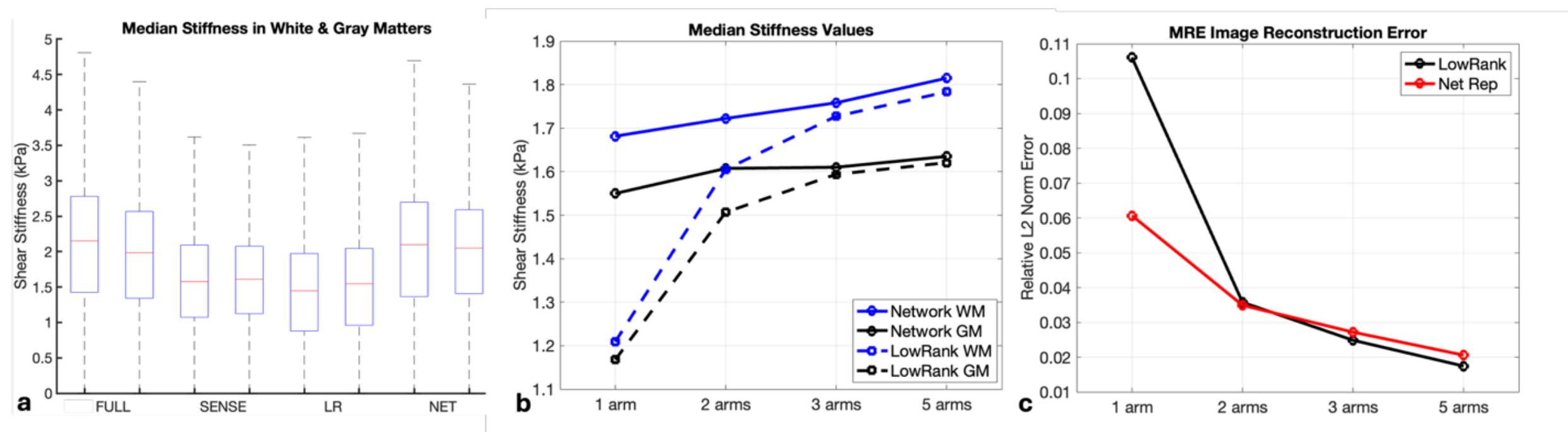

Figure 5. Quantitative comparison of estimated stiffness and reconstructed MRE images across different methods and undersampling factors. a) Stiffness estimated from the spin-echo dataset using SENSE, low-rank and the proposed DL methods from a single spiral arm (left bar: white matter; right bar: gray matter). b) Estimated median stiffness and c) relative L2 norm errors of the reconstructed MRE image from the 3D gradient-echo dataset using the low-rank and the proposed DL methods with varying number of spiral arms.

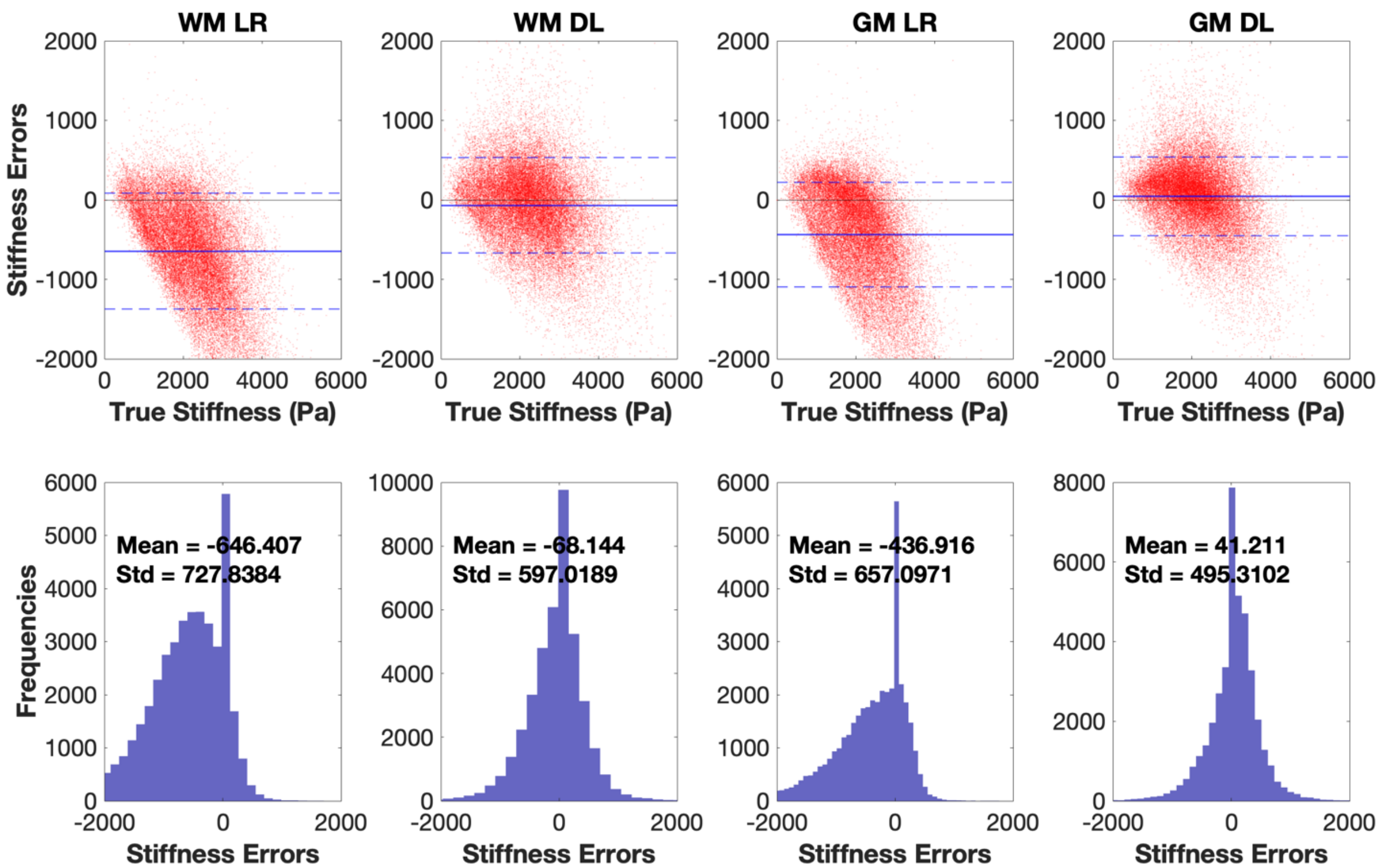

Figure 6. Quantitative comparison of stiffness estimation in white matter and gray matter from the spin-echo dataset using low-rank (LR) and proposed (DL) methods. (Top) plots of stiffness errors versus the "true" stiffness values from fully sampled data and (Bottom) histograms of stiffness errors.

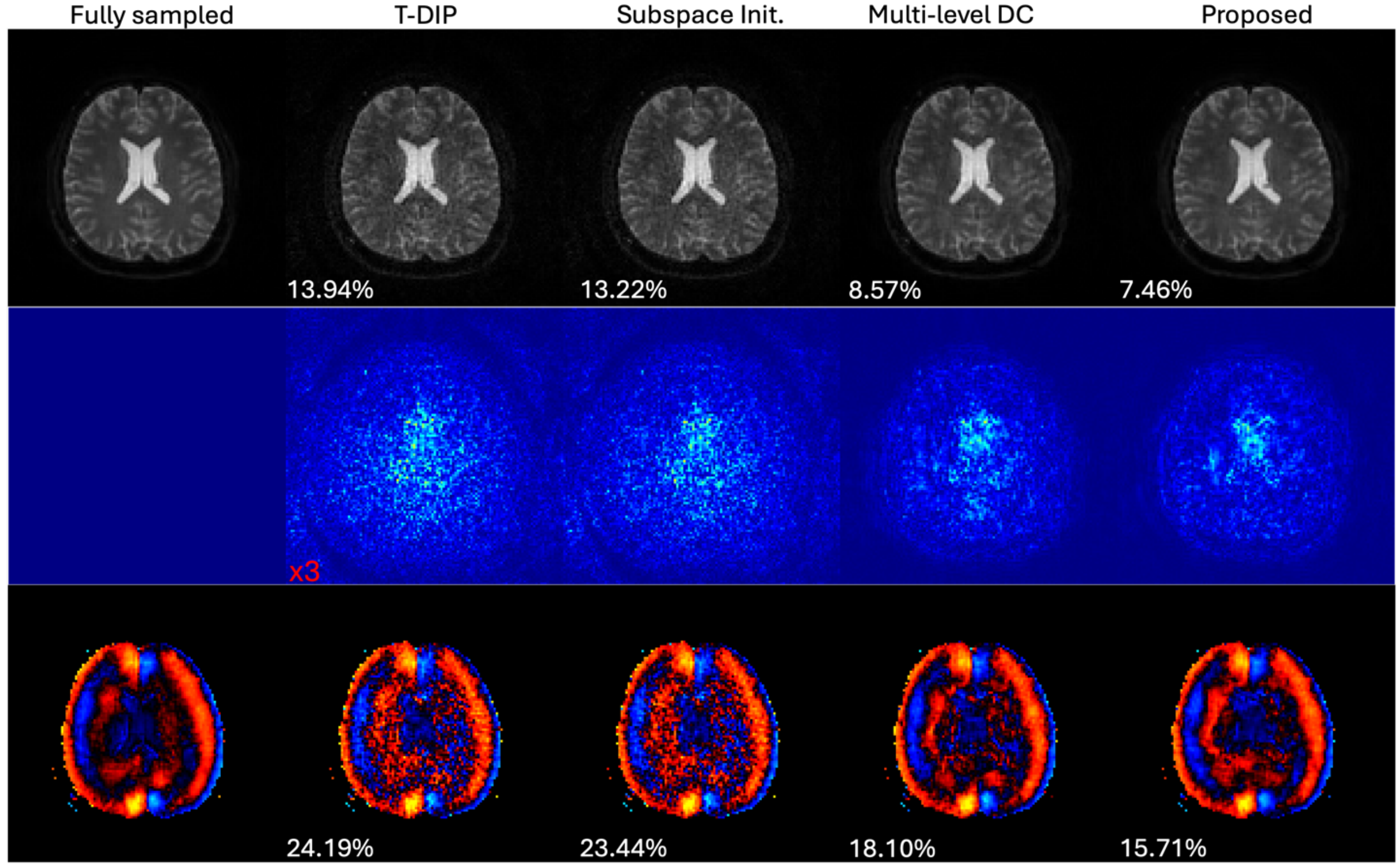

Figure 7. Ablation study using the spin echo data set comparing Time-DIP method with random initialization, temporal subspace initialization, along with the addition of multi-level data consistency and the elastography-specific constrains. Relative L2 norm errors of the anatomical and wave reconstructions were shown at the lower left corner.

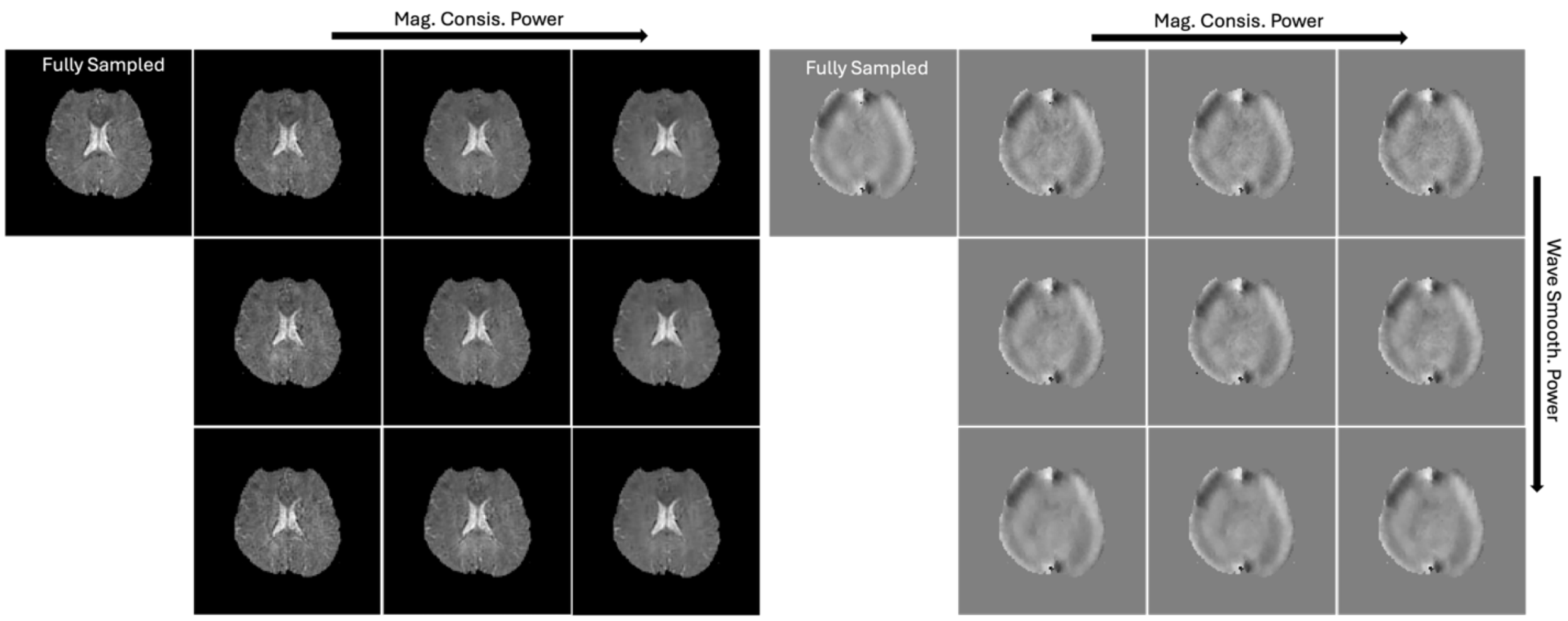

Figure 8. The distinct roles of magnitude similarity and displacement smoothness constraints under varying combinations of regularization parameters (i.e., weak, moderate and strong).